# Generation of Anti-Stokes Fluorescence in a Strongly Coupled Organic Semiconductor Microcavity

Kyriacos Georgiou[1], Rahul Jayaprakash[1], Alexis Askitopoulos[2,3], David M. Coles[1], Pavlos G. Lagoudakis[2,3]* and David G. Lidzey[1]*

[1]Department of Physics and Astronomy, The University of Sheffield, Hicks Building, Hounsfield Road, Sheffield, S3 7RH, United Kingdom.
[2]Department of Physics and Astronomy, University of Southampton, Southampton, SO17 1BJ, United Kingdom.
[3]Skolkovo Institute of Science and Technology, Moscow 143026, Russia.

**Abstract**

We explore the generation of anti-Stokes fluorescence from strongly coupled organic dye microcavities following resonant ground-state excitation. We observe polariton emission along the lower polariton branch, with our results indicating that this process involves a return to the exciton reservoir and the absorption of thermal energy from molecules in a vibrationally excited ground-state. We speculate that the generation of a population of 'hot' polaritons is enhanced by the fact that the cavity supresses the emission of Stokes-shifted fluorescence, as it is energetically located below the cut-off frequency of the cavity.

Over the past two decades, optical microcavities have been used to explore the physics of optical-coupling between confined-photons and molecular semiconductors.[1–7] Optical microcavities are structures that consist of two mirrors placed in close proximity that are separated by a layer of an optically active material. The mirrors confine the local electromagnetic field, and when it is resonant with the excitonic transition of the semiconductor, a reversible exchange of energy can occur between light and matter. In the strong-coupling regime, this results in the formation of quasi-particles termed cavity-polaritons that are coherent superposition of excitons and photons and have properties of both.[8] By coupling molecular materials with light, it is possible to engineer fascinating processes, such as polariton condensation,[9–11] superfluidity,[6] changes in the rate of chemical reactions[12,13] and a modification of energy transfer pathways.[14–17]

The first microscopic models used to describe the strong-coupling of molecules to light treated the molecules in the cavity as an ideal two-level system and neglected rotational or vibrational degrees of freedom.[18] Recently however, there has been growing interest in understanding the role that the vibrational landscape plays in modifying basic processes in strongly-coupled microcavities. Early experimental work on the resonant excitation of organic strongly-coupled cavities demonstrated an enhancement of Stokes and anti-Stokes Raman scattering process when the energetic separation between initial and final states were degenerate with a Raman mode of the molecular material within the cavity.[19] Follow-on studies established that the (Raman-active) vibrational modes of a molecule were able to assist the relaxation of uncoupled excitons from the 'exciton reservoir' to polaritonic states along the lower polariton branch (LPB) by acting as a sink for excess energy.[20,21]

In inorganic semiconductor microcavities a number of authors have explored the interactions between cavity-polaritons and acoustic and LO phonons.[22–29] For example, anti-Stokes fluorescence (ASF) was observed in strongly-coupled optical microcavities containing ZnSe quantum wells[30] following resonant excitation of lower-branch polaritons,[31] with ASF used to achieve laser cooling of semiconductors.[32–34] In such microcavities,[31] the significant excitonic component of the cavity-polaritons was responsible for the interactions with thermal phonons, making them an important platform in which to study laser-cooling effects in semiconducting materials. Indeed, by exciting the LPB at $k_{//}=0$, the generation of fluorescence at lower energy is forbidden due to the near parabolic dispersion of cavity-polaritons.

In organic-semiconductor based microcavities, a growing interest in the interactions between cavity-polaritons and rovibrational modes of molecular materials has stimulated much theoretical research.[35–41] For example, it has been predicted that electronic and nuclear degrees of freedom become decoupled in the strong coupling regime; an effect that reduces disorder and enhances the exciton coherence length.[36] Furthermore, models that include rotational and vibrational degrees of freedom indicate that dark states within a microcavity (corresponding to uncoupled excitons in to a two-level system) can also be affected by strong coupling.[37]

In this paper, we provide evidence for the interaction between cavity-polaritons with the local thermal bath in a strongly-coupled microcavity containing a film of the fluorescent molecular dye bromine-substituted boron-dipyrromethene (BODIPY-Br). We use optical spectroscopy to show that ASF can be generated by exciting the LPB around $k_{//} = 0$, and perform temperature dependent measurements that show that by reducing the overall thermal energy available in the system we can control the intensity of polariton ASF, suggesting a thermally assisted process. We utilise fluorescence lifetime measurements to demonstrate that the polariton ASF process involves a 'return' of energy to the exciton reservoir. We find that the overall integrated population of 'hot' polaritons in the resonantly excited strongly-coupled cavity (corrected for the number of photons absorbed) is around 3 times larger than the relative number of ASF photons generated from a control (non-cavity) film of BODIPY-Br. We speculate that the cavity suppresses the emission of Stokes-shifted luminescence as it is located energetically below the cut-off frequency of the cavity.

The chemical structure of BODIPY-Br, together with its absorption and fluorescence are shown in Figure 1a. Here, the peak in absorption and fluorescence observed at around 530 and 550 nm respectively corresponds to the 0-0 electronic transition of the BODIPY-Br monomer, with the 20 nm difference between the wavelength of the absorption and emission maxima occurring as a result of a Stokes shift.[42] The absorption shoulder at 503 nm corresponds to a 0-1 vibrational transition. The shoulder in PL emission at 586 nm has a more complex origin, and corresponds to emission from the 0-1 vibrational transition together with emission from excited state dimers (excimers).[43]

To fabricate microcavities, an optically inert polymer matrix (polystyrene - PS) was doped with BODIPY-Br at a dye/polymer concentration of 20%/80% by mass and spin-cast onto a 200 nm thick silver mirror forming a film having a thickness of 260 nm. A film of polyvinyl alcohol (PVA) was spin-cast on top of the BODIPY-Br layer to tune the cavity length, and a semi-transparent silver mirror was then thermally evaporated onto the organic bilayer having a thickness of 32 nm. The resultant λ-cavity had a Q-factor of ~50. A schematic of a typical cavity is shown in Figure 1b. We have recently shown that the BODIPY-Br 0-0 monomer excitonic transition can undergo strong coupling [44], and when incorporated into higher Q-factor cavities based on two distributed Bragg reflectors (DBRs), such cavities demonstrate polariton-condensation.[11]

Using white light reflectivity measurements we find that the cavities we have constructed are in the strong-coupling regime. Here, all measurements were performed through the semi-transparent top mirror, allowing light to be coupled into and out of the cavity. We plot the cavity reflectivity as a function of external viewing angle in Figure 1c. We find that three dispersive optical features are detected, which can be described using a three-level coupled-oscillator model as plotted in Figure 1c. From this, we determine a Rabi splitting of 210 and 80 meV between the cavity photon and the transitions $X_{0-0}$ and $X_{0-1}$ respectively. This suggests that the cavity photon (C) undergoes strong-coupling with the 0-0 electronic transition of BODIPY-Br, with the 0-1 vibronic transition of BODIPY-Br remaining in an intermediate-coupling regime.[45] Our model also indicates that the cavity is negatively detuned, with the energetic separation between $X_{0-0}$ and the cavity photon at $\theta = 0°$ being 85 meV. In the Supporting Information, we plot the Hopfield coefficients for the three polariton-branches (see Supporting Information Figure S1). From this analysis, we determine that the bottom of the LPB is composed of 69% cavity photon, with the remainder being a mixture of $X_{0-0}$ and $X_{0-1}$ (29% and 2% respectively).

We have explored the emission of fluorescence following wavelength-dependent excitation of strongly coupled cavities and control (non-cavity) films of BODIPY-Br / PS as a function of temperature. Room temperature emission from a *control film* of BODIPY-Br / PS performed as a function of excitation wavelength is shown in Figure 2. Here, the excitation wavelength was varied between 500 nm (corresponding to a spectral-region with high direct absorption) and 580 nm (a 'sub band-gap' spectral-region corresponding to very low direct absorption). The vertical dashed line at 550 nm corresponds to the peak emission wavelength of the $X_{0-0}$ electronic transition of BODIPY-Br. It can be seen that the BODIPY-Br undergoes clear frequency up-conversion, with ASF emitted at an energy more than 160 meV above the laser excitation. This effect is enhanced at excitation wavelengths closer to the edge of the BODIPY-Br absorption band (560 nm) with the intensity reducing with increasing excitation wavelength.

We have also characterised ASF emission from control films as a function of temperature, with excitation performed at different wavelengths. Typical data is shown in Figures 3a and b recorded at temperatures between 80 and 300K following excitation at 548 and 558 nm respectively. We find that as the temperature is reduced, the intensity of ASF is reduced, with its spectral distribution shifting to longer wavelengths. Figure 3c plots emission spectra recorded from a BODIPY-Br / PS control film as a function of temperature following excitation at 500 nm. Here, we again observe a redistribution of PL emission to longer wavelengths as the temperature is reduced; an effect that we have previously shown results from emission mainly originating from isolated BODIPY-Br monomers at room temperature and from intermolecular excimers at 80 K.[43]

Previous work has explored ASF in a number of organic materials [33,34] and it has been proposed that optical-excitation excites a subset of molecules that are initially in a vibrationally-excited ground-state to a higher lying $S_1$ state. On relaxation, the molecules return to their electronic (0-0) ground state, with the luminescence emitted having an energy greater than that of the incident photon. Following the approach outlined in refs [33,34], we calculate an activation energy for the ASF emission. Here, measurements were made on a control film, with emission intensity measured at 547 nm as a function of temperature (I(T)) following photo-excitation at 568 nm. Typical data is plotted in Figure 3d, which we fit using an Arrhenius equation of the form ($I(T) = Ce^{-E_A/k_B T}$), where $E_A$ is an activation energy of around 27 meV. This activation energy has previously been associated with the energy difference between the energy of the excitation-photons and the 0-0 absorption transition. However this is clearly not the case here, as the activation energy (27 meV) is much smaller that the energy-separation between 0-0 transition and laser excitation energy (in this case 156 meV). Indeed, even when we excite the film at shorter wavelengths, we only observe a small change in the activation energy (see Supporting Information Figure S3). We note that in experiments on Rhodamine 101 in solution,[34] the activation energy and the calculated energy difference between the excitation and the 0-0 transition were found to be in good agreement. However measurements on Rhodamine B dispersed into a solid film[33] indicated that the activation energy was 62 meV (500 cm$^{-1}$) smaller than its expected value, with this discrepancy ascribed to solvent effects. In our work on BODIPY-Br, we find a clear discrepancy of around 120 meV between the measured activation energy and the energy difference between excitation and the 0-0 transition energies. The reason for this discrepancy is not clear, however we suspect it may result from the fact that BODIPY-Br emission spectra has a significant temperature dependence that results from migration of energy to intermolecular excimer states (see temperature dependent PL spectra shown in Figure 3a-c). Nevertheless the similarity between our findings and those in ref[33] suggests that the ASF emission generated in BODIPY-Br / PS control films also results from the direct photo-excitation of molecules that are in a vibrationally excited ground-state.

In the following section, we analyse ASF emission from microcavities containing strongly coupled BODIPY-Br. Figure 4 shows typical data from a microcavity in which the LPB was excited resonantly at 572 nm; a wavelength corresponding to the bottom of the LPB, with temperature progressively tuned between 77 and 292 K. Here, a small blue-shift of the polariton dispersion by around 1 nm is observed over this temperature range; an effect that most likely results from a small thermally-induced contraction of the cavity optical path-length. The dashed line in Figure 4 shows the dispersion of the LPB as fitted using the measured white light reflectivity (Figure 1c), with the colour scale quantifying the intensity of polariton emission. It can be seen that at 292 K, significant luminescence emission is observed along the LPB at energies up to 95 meV above that of the laser-excitation (which can be identified by the horizontal 'red' band); a result consistent with ASF. We find that as the cavity temperature is progressively decreased, emission is still observed along the LPB, however, the intensity of ASF reduces with temperature and at low temperatures emission is concentrated around the bottom of the LPB. Note, that while it was possible to observe ASF that extended 160 meV above the energy of the laser in the non-cavity control films, it was not possible to observe ASF at such large energies above the excitation laser in the strongly-coupled cavity shown in Fig 1c. The reason for this is that the dispersion of the LPB flattens considerably as it approaches the energy of the exciton, and thus such large energy-separations are inaccessible. We have however observed polariton ASF in cavities having a larger negative detuning, permitting a greater energy separation (~135 meV) to be achieved between excitation and ASF emission (see

Figure S5 in Supporting Information). It is clear however that the initially excited states in such negatively-detuned cavities are highly photon-like in character (having a total photon fraction of 97%).

To understand the ASF process in more detail, Figure 5a plots the intensity of polariton ASF at an angle of 30º (corresponding to a wavelength of 555 nm) as a function of temperature, following excitation at 572 nm. We again use an Arrhenius fit to describe the temperature-dependent intensity, and extract an activation energy of around 27 meV; a result in close agreement with that determined from the non-cavity BODIPY / PS control film (see Figure 3d).

In Figure 5b we plot the normalised polariton population along the LPB as a function of external angle. Here, polariton population ($P_{LPB}$) is calculated using $P_{LPB}(\theta) = I(\theta)/\alpha_c(\theta)$, where $I$ is the intensity of luminescence and $\alpha_c$ is the photon fraction of the LPB at angle $\theta$, (calculated using a 3-level coupled oscillator model - see Supporting Information). It can be seen in Figure 5a that while the intensity of ASF along the LPB reduces as a function of temperature, the relative distribution of polariton population along the LPB does not change (see Figure 5b). We can in fact determine an effective temperature of the polaritons along the LPB by performing a Boltzmann fit as shown in the inset of Figure 5b. We find that the effective temperature of the polaritons to be around 590 K. The fact that this effective temperature is approximately constant as a function of cavity temperature suggests that the polariton population does not reach thermal equilibrium; a result consistent with the very fast radiative lifetime of polaritons in a low Q-factor cavity.

Our measurements demonstrate therefore that following excitation of polaritons at the bottom of the LPB, a process occurs in which states at higher energy along the LPB are populated. The apparent similarity between the activation energy for ASF emission in both the cavity and control films indicates that we should expect a similar underlying mechanism. To explore this in more detail, we have measured the dynamics of polariton emission from states along the LPB using a streak camera following both resonant and non-resonant excitation. Typical data is shown in Figure 6a; photoluminescence is recorded from the cavity at around 556 nm (corresponding to an angle $\theta$ = 28°), where the cavity is either excited resonantly at 570 nm, or non-resonantly at 400 nm. We find the decay dynamics are in both cases bi-exponential, with polaritons excited resonantly decaying with a lifetime of 57 and 286 ps and relative amplitudes of 0.645 and 0.355 respectively (fit to the data having an adjusted R-squared value of 0.999). Polaritons excited non-resonantly instead decay with a lifetime of 71 and 518 ps and relative amplitudes of 0.59 and 0.41 respectively (adjusted R-squared fit value of 0.997). Our results indicate therefore that the ASF polariton decay lifetime is a factor of approximately 2 shorter than emission generated following non-resonant excitation (where the initial photo-generated states are excitons). We believe that the relative similarity between the recorded lifetimes indicates that the ASF process must involve interaction with states in the exciton reservoir (the radiative lifetime of polaritons in this low Q-factor cavity is estimated to be around 30 fs). The reason for this apparent reduction of lifetime most likely indicates that the ASF process is subject to additional non-radiative losses – a conclusion consistent with the higher intensity of luminescence emitted when the cavity is pumped non-resonantly compared to resonantly generated ASF emission.

We can further evidence the importance of uncoupled-excitons in mediating the polariton ASF process by modifying the energetic distribution of states within the exciton reservoir. Here, a molecular 'acceptor' species was introduced at low concentration into a BODIPY-Br / PS film, with the acceptor positioned at an energy below that of the LPB. The acceptor molecules used were based on a related molecular dye BODIPY-R that has a reduced electronic energy-gap as a result of chemical modification. This molecule acts as a low-energy acceptor, with control measurements on blends of BODIPY-Br / BODIPY-R (20% / 0.75% by mass) dispersed in polystyrene indicating almost complete energy transfer (> 99%) from BODIPY-Br to BODIPY-R by dipole-dipole coupling. This efficient dipole-dipole energy-transfer can be seen in Supporting Information Figure S6a, where we plot absorption and PL emission from a blend BODIPY-Br / BODIPY-R / PS film, with excitation performed at 473 nm. Here, PL is dominated by emission from BODIPY-R, with emission from BODIPY-Br being almost totally absent. In Figure S6b, we show emission from the same blend, where excitation was performed at 575 nm. Again, emission is dominated by BODIPY-R, with no ASF from BODIPY-Br detected.

We have explored the photophysics of such BODIPY-Br / BODIPY-R / PS films placed within a microcavity (with the cavity having a very similar mode-structure as the BODIPY-Br microcavity described above). Here, the BODIPY-Br was strongly-coupled, however by using the BODIPY-R acceptor at low concentration, we ensured that it could not undergo strong coupling. We find that when such a BODIPY-Br / BODIPY-R / PS blend was used as the active layer in a microcavity *no detectable ASF* was emitted from the cavity on resonant excitation of the LPB at 572 nm. This indicates that any polaritons that return to the exciton reservoir undergo energy transfer to weakly-coupled BODIPY-R exciton states positioned at lower energy rather than scattering back to higher energy polariton states. We conclude therefore that the ASF process evidenced in strongly coupled BODIPY-Br microcavities is mediated by an exchange with the exciton reservoir and does not involve a direct polariton to polariton scattering mechanism along the lower polariton branch.

We can gain additional insight into the ASF process, through a quantitative comparison of the angular dependence of ASF emission detected from a cavity and a non-cavity control film. This is shown in Figure 6b, where we plot the relative population of cavity polaritons ($P_{LPB}(\theta)$) as a function of angle along the LPB at room-temperature, with data shown for either resonant (572 nm) or non-resonant (500 nm) excitation. On the same figure, we plot the intensity of Stokes and anti-Stokes emission as a function of angle recorded from a control (non-cavity) BODIPY-Br film on excitation at 570 nm (corresponding to sub-bandgap excitation). Note, that we do not apply any correction to the control-film emission intensity, and assume there is a one-to-one correspondence between photoluminescence intensity and the exciton-population. However, we have measured the relative number of excitation photons that were absorbed by both the cavity and the control film following resonant and non-resonant excitation, and applied an appropriate correction in both cases to allow us compare relative populations. For completeness, we also plot the intensity of emission as a function of angle generated from the control film following excitation at 500 nm.

It can be seen that when the cavity and the control film are excited at 500 nm (green and magenta lines respectively), we find that a very similar population of states are created. This indicates that in both the cavity and the control film, the initially generated 'hot' excitons undergo a similar relaxation process. In the cavity, such relaxed excitons form a reservoir of states that then populate states along the LPB by radiatively pumping its photonic component with high efficiency; a result in accord with our previous findings.[44] It can be also seen that the angular emission from the cavity, when excited at 500 nm

(green line) is much higher than that following excitation at 572 nm (black line). This indicates that the ASF process in the cavity is subject to additional *unknown* non-radiative losses as compared to a film excited at significantly shorter wavelengths; a result consistent with the enhanced decay-rate of ASF emission shown in Fig 6a.

When instead the cavity and control film are excited at 572 and 570 nm (corresponding to the bottom of the LPB and a sub-band gap region of BODIPY-Br respectively) we find that the total angle-integrated polariton population that results in ASF emission from the cavity is 3 times larger than the relative number of ASF photons emitted by the film (black and red lines respectively). When however we also include Stokes shifted emission (blue line) from the control film (i.e. emission at wavelengths greater than the excitation laser at 570 nm), we find that the angle-integrated ASF polariton population from the LPB and the overall angle-integrated fluorescence from the control film (Stokes and anti-Stokes) are quantitatively similar.

We offer the following picture to explain this effect (see schematic shown in Figure 7a-c). On excitation of the non-cavity control film at an energy $E_i$ below the energy-gap (i.e. at 570 nm), we select molecules in a vibrationally excited ground-state ($\Delta E$) and raise such molecules to the zero-phonon (1-0) excited state. On return to their ground-state, the luminescence from such molecules includes the initial ground-state vibrational-energy ($E_i + \Delta E$), and thus part of the emitted luminescence occurs at shorter wavelengths (corresponding to ASF – see Figure 7a). Note that Stokes emission is also possible, where the molecule returns to a vibrationally excited ground state ($E_i - \Delta E$) (see Figure 7b).

On resonant excitation of the cavity at the bottom of the LPB, polaritons are created, that we speculate can exchange energy with weakly-coupled electronic states in the exciton reservoir. Here, such weakly-coupled states most likely correspond to molecules in a vibrationally excited ground-state ($\Delta E$), with this excitation process probably involving the direct absorption of the photon-component of the polariton. This process then creates a weakly-coupled reservoir exciton. Again, we assume the initially excited polariton state has an energy $E_i$ (see Figure 7c), with the reservoir-exciton thus created having an energy of ($E_i + \Delta E$). This exciton is then able to radiatively pump polaritons along the LPB[44] having an energy ($E_i + \Delta E$).

This simple model can also qualitatively explain the relative enhancement of 'hot' polariton population corresponding to ASF emission observed when resonantly exciting the cavity as compared to the control film. If the weakly-coupled exciton created in the reservoir instead undergoes Stokes emission, the photon emitted will have an energy ($E_i$ - $\Delta E$) and will thus be below the cut-off frequency of the cavity (see Figure 7c). For this reason, we expect Stokes-shifted emission to be relatively suppressed (via the Fermi's Golden Rule) through the relatively low density of optical states at such energies. This results in a re-distribution of emission, where anti-Stokes emission becomes the dominant decay mechanism. Note, however, that this simple picture ignores the fact that there may well be a high density of waveguided modes that exist at high wavevector into which Stokes shifted emission could couple. At the moment, the relative importance of such waveguide modes is not understood, and further detailed modelling is required to produce a more complete picture of the total density of optical states into which molecules could emit Stokes or anti-Stokes radiation.

To conclude, we have provided direct experimental evidence of anti-Stokes fluorescence (ASF) from a strongly-coupled organic-semiconductor microcavity that is resonantly excited at the bottom of the lower polariton branch (LPB). We have characterised the ASF as a function of temperature, and have compared it with that emitted by a control (non-cavity) film composed of the same organic semiconductor that is excited at a very similar wavelength. We find that the angle-integrated 'hot' polariton population from the cavity LPB is around 3 times larger than that corresponding to ASF emission by the control film. By measurement of the dynamics of ASF emission, we conclude that this process involves a return of the photo-generated polaritons to a weakly-coupled exciton reservoir. We propose a picture in which the polariton states return to the reservoir by exciting molecules that are in a vibrationally excited ground-state. The excitons created are then able to optically pump polariton states above the bottom of the LPB, with the excess energy provided by the initial ground-state vibrational energy of the molecule. The relative enhancement of ASF states results from the fact that Stokes-shifted fluorescence is suppressed as its energy is below the cut-off frequency of the cavity.

Our measurements contribute to the growing understanding of the importance of the role played by vibrational modes in underpinning basic processes in strongly-coupled organic semiconductor microcavities,[20,21,48,49,35–40,46,47] and it will clearly be important to include such processes when constructing a full model of organic exciton-polariton condensation. We also expect that structures similar to those explored may permit a study of the physics of laser cooling. Critically, the ability to undergo laser cooling at room temperature distinguishes organic-based systems from inorganic-semiconductor based microcavities that (in many cases) only demonstrate polaritonic effects at cryogenic temperatures. Note however that we have not yet evidenced laser cooling in the cavities or films studied. Clearly, to create a cooling effect, the energy flux entering the cavity must be smaller than that leaving the cavity, however this is unlikely to be the case at present. Indeed the BODIPY-Br films used have a relatively low (~5%) photoluminescence quantum efficiency (PLQE) as a result of molecular aggregation[43] and thus a significant quantity of the absorbed laser is expected to be dissipated as heat. Nevertheless, we expect that strongly-coupled cavities containing molecular dyes having a significantly higher PLQE may allow laser cooling effects to be explored at room temperature.

## METHODS

**Sample preparation.** Polystyrene (PS) having an average molecular weight ($M_w$) of ~192 000 was dissolved in toluene at a concentration of 35 mg mL$^{-1}$. Poly-vinyl alcohol (PVA) was dissolved in DI water at a concentration of 35 mg mL$^{-1}$. BODIPY-Br was added into the PS/toluene solution at a concentration of 20% by mass. Thin control films were spin-cast on quartz-coated glass substrates. To fabricate a microcavity, a 200 nm thick silver mirror was thermally evaporated on a quartz-coated glass substrate using an Auto 360 Edwards thermal evaporator. BODIPY-Br/PS and PVA layers were then spin-cast on top of the silver mirror. The thickness of the organic layers was measured using a Bruker DektakXT profilometer and the spin-speed of the spin coater was adjusted to achieve the desired thickness. The top semi-transparent silver mirror was then thermally evaporated to have a thickness of 32 nm.

**Angular white-light reflectivity.** White-light reflectivity measurements were performed through the top semi-transparent mirror using light from a fibre-coupled Halogen-Deuterium lamp that was imaged using two lenses onto the cavity surface, with the reflected light collected using a second pair of lenses and imaged into an Andor Shamrock CCD spectrometer. Both pairs of lenses were mounted on motorised optical rails fixed to a common rotation stage, with reflectivity measurements performed as a function of angle in steps of 1º.

**Fluorescence measurements.** The microcavities and control film were mounted in a liquid $N_2$ cryostat, with excitation performed using a Fianium supercontinuum white light laser emitting 6 ps pulses at a repetition rate of 40 MHz. Laser light was first filtered using a SPEX 270M monochromator having a bandwidth of 1.7 nm, allowing the excitation wavelength to be tuned over the range of 500 to 600 nm. The filtered laser light was then coupled into an optical fibre and imaged onto the sample surface. The resultant emission was collected at external angles between 18º and 70º using the same optical rail setup used in reflectivity measurements. All data is corrected for the Lambertian response of the system. All microcavities were excited at normal incidence ($k_{//}=0$).

**Streak camera measurements.** For non-resonant excitation measurements (400 nm) a frequency-doubled Ti:sapphire regenerative amplifier was used with a 180 fs pulse length, while for the quasi resonant excitation (570 nm) an optical parametric oscillator was used with a spectrally filtered bandwidth of 2nm and 550 fs pulse length. Both measurements were performed with a 100 KHz repetition rate, with the transient emission dynamics recorded using a streak camera with a temporal resolution of 2 ps.

## AUTHOR INFORMATION

**Corresponding Authors**

*E-mail: d.g.lidzey@sheffield.ac.uk and pavlos.lagoudakis@soton.ac.uk

**ORCID**

Kyriacos Georgiou: 0000-0001-5744-7127

David M. Coles: 0000-0003-4808-6395

Pavlos G. Lagoudakis: 0000-0002-3557-5299

David G. Lidzey: 0000-0002-8558-1160

**Author Contributions**

K.G., P.G.L. and D.G.L. designed the experiments. K.G. fabricated the samples. K.G., R.J. and D.M.C. performed the reflectivity and steady state fluorescence measurements. K.G. and A.A. performed the transient emission measurements.  K.G., R.J., A.A. and D.M.C. analysed the data. K.G., P.G.L and D.G.L wrote the manuscript. All authors discussed the results and contributed to the manuscript.

**Notes**

The authors declare no competing financial interest.

## ACKNOWLEDGEMENTS

We thank Marco Cavazzini and Francesco Galleotti (Milano) for the synthesis of the BODIPY-Br used in these experiments. We thank Zhen Shen and Lizhi Gai (Nanjing University) for the synthesis of the BODIPY-R. We are grateful to Paolo Michetti (Dresden University), Jenny Clark and Andrew J. Musser (University of Sheffield) for fruitful discussions. We thank the U.K. EPSRC for funding this work via the Programme Grant 'Hybrid Polaritonics' (EP/M025330/1) and for funding a Ph.D. scholarship for K.G. through a DTG allocation.

## REFERENCES

(1)  D. G. Lidzey; Bradley, D. D. C.; Skolnick, M. S.; Virgili, T.; Walker, S.; Whittaker, D. M. Strong Exciton-Photon Coupling in an Organic Semiconductor Microcavity. *Lett. to Nat.* **1998**, *395*, 53–55.

(2) Lidzey, D. G.; Bradley, D. D. C.; Virgili, T.; Armitage, a.; Skolnick, M. S.; Walker, S. Room Temperature Polariton Emission from Strongly Coupled Organic Semiconductor Microcavities. *Phys. Rev. Lett.* **1999**, *82* (16), 3316–3319.

(3) Lidzey, D. G.; Bradley, D. D. C.; Armitage, A.; Walker, S.; Skolnick, M. S. Photon-Mediated Hybridization of Frenkel Excitons in Organic Semiconductor Microcavities. *Science* **2000**, *288* (5471), 1620–1623.

(4) Kéna-cohen, S.; Forrest, S. R. Room-Temperature Polariton Lasing in an Organic Single-Crystal Microcavity. *Nat. Photonics* **2010**, *4*, 371–375.

(5) Lerario, G.; Ballarini, D.; Fieramosca, A.; Cannavale, A.; Genco, A.; Mangione, F.; Gambino, S.; Dominici, L.; De Giorgi, M.; Gigli, G.; Sanvitto, D. High-Speed Flow of Interacting Organic Polaritons. *Light Sci. Appl.* **2017**, *6*, e16212.

(6) Lerario, G.; Fieramosca, A.; Barachati, F.; Ballarini, D.; Daskalakis, K. S.; Dominici, L.; De Giorgi, M.; Maier, S. A.; Gigli, G.; Kéna-Cohen, S.; Sanvitto, D. Room-Temperature Superfluidity in a Polariton Condensate. *Nat. Phys.* **2017**, *13*, 837–841.

(7) Coles, D.; Flatten, L. C.; Sydney, T.; Hounslow, E.; Saikin, S. K.; Aspuru-Guzik, A.; Vedral, V.; Tang, J. K. H.; Taylor, R. A.; Smith, J. M.; Lidzey, D. G. A Nanophotonic Structure Containing Living Photosynthetic Bacteria. *Small* **2017**, *13*, 1701777.

(8) Weisbuch, C.; Nishioka, M.; Ishikawa, A.; Arakawa, Y. Observation of the Coupled Exciton-Photon Mode Splitting in a Semiconductor Quantum Microcavity. *Phys. Rev. Lett.* **1992**, *69* (23), 3314–3317.

(9) Plumhof, J. D.; Stoeferle, T.; Mai, L.; Scherf, U.; Mahrt, R. F. Room-Temperature Bose-Einstein Condensation of Cavity Exciton-Polariton in a Polymer. *Nat. Mater.* **2014**, *13*, 247–252.

(10) Daskalakis, K. S.; Maier, S. a; Murray, R.; Kéna-cohen, S. Nonlinear Interactions in an Organic Polariton Condensate. *Nat. Mater.* **2014**, *13*, 271–278.

(11) Cookson, T.; Georgiou, K.; Zasedatelev, A.; Grant, R. T.; Virgili, T.; Cavazzini, M.; Galeotti, F.; Clark, C.; Berloff, N. G.; Lidzey, D. G.; Lagoudakis, P. G. A Yellow Polariton Condensate in a Dye Filled Microcavity. *Adv. Opt. Mater.* **2017**, *5* (18), 1700203.

(12) Hutchison, J. A.; Schwartz, T.; Genet, C.; Devaux, E.; Ebbesen, T. W. Modifying Chemical Landscapes by Coupling to Vacuum Fields. *Angew. Chemie - Int. Ed.* **2012**, *51*, 1592–1596.

(13) Hutchison, J. A.; Liscio, A.; Schwartz, T.; Canaguier-Durand, A.; Genet, C.; Palermo, V.; Samorì, P.; Ebbesen, T. W. Tuning the Work-Function via Strong Coupling. *Adv. Mater.* **2013**, *25*, 2481–2485.

(14) Coles, D. M.; Somaschi, N.; Michetti, P.; Clark, C.; Lagoudakis, P. G.; Savvidis, P. G.; Lidzey, D. G. Polariton-Mediated Energy Transfer between Organic Dyes in a Strongly Coupled Optical Microcavity. *Nat. Mater.* **2014**, *13*, 712–719.

(15) Zhong, X.; Chervy, T.; Wang, S.; George, J.; Thomas, A.; Hutchison, J. A.; Devaux, E.; Genet, C.; Ebbesen, T. W. Non-Radiative Energy Transfer Mediated by Hybrid Light-Matter States. *Angew. Chemie - Int. Ed.* **2016**, *55*, 6202–6206.

(16) Zhong, X.; Chervy, T.; Zhang, L.; Thomas, A.; George, J.; Genet, C.; Hutchison, J.; Ebbesen, T. W. Energy Transfer between Spatially Separated Entangled Molecules. *Angew. Chemie Int. Ed.* **2017**, *56*, 9034–9038.

(17) Georgiou, K.; Michetti, P.; Gai, L.; Cavazzini, M.; Shen, Z.; Lidzey, D. G. Control over Energy Transfer between Fluorescent BODIPY Dyes in a Strongly-Coupled Microcavity. *ACS Photonics* **2018**, *5* (1), 258–266.

(18) Michetti, P.; Mazza, L.; Rocca, G. C. La. Strongly Coupled Organic Microcavities. In *Organic Nanophotonics, Nano-Optics and Nanophotonics*; Sheng, Y., Ed.; Springer: Berlin, 2015; pp 39–68.

(19) Tartakovskii, A.; Emam-Ismail, M.; Lidzey, D. G.; Skolnick, M. S.; Bradley, D. D. C.; Walker, S.; Agranovich, V. M. Raman Scattering in Strongly Coupled Organic Semiconductor Microcavities. *Phys. Rev. B* **2001**, *63* (12), 121302.

(20) Coles, D. M.; Michetti, P.; Clark, C.; Tsoi, W. C.; Adawi, A. M.; Kim, J. S.; Lidzey, D. G. Vibrationally Assisted Polariton-Relaxation Processes in Strongly Coupled Organic-Semiconductor Microcavities. *Adv. Funct. Mater.* **2011**, *21* (19), 3691–3696.

(21) Mazza, L.; Kéna-Cohen, S.; Michetti, P.; La Rocca, G. C. Microscopic Theory of Polariton Lasing via Vibronically Assisted Scattering. *Phys. Rev. B* **2013**, *88* (7), 075321.

(22) Li, W.; Gao, M.; Zhang, X.; Liu, D.; Peng, L. M.; Xie, S. Microphotoluminescence Study of Exciton Polaritons Guided in ZnO Nanorods. *Appl. Phys. Lett.* **2009**, *95* (17), 173109.

(23) Bœuf, F.; André, R.; Romestain, R.; Dang, L. S.; Péronne, E.; Lampin, J. F.; Hulin, D.; Alexandrou, A. Evidence of Polariton Stimulation in Semiconductor Microcavities. *Phys. Rev. B* **2000**, *62* (4), R2279–R2282.

(24) Delteil, A.; Vasanelli, A.; Jouy, P.; Barate, D.; Moreno, J. C.; Teissier, R.; Baranov, A. N.; Sirtori, C. Optical Phonon Scattering of Cavity Polaritons in an Electroluminescent Device. *Phys. Rev. B* **2011**, *83* (8), 081404.

(25) Müller, M.; Bleuse, J.; Andre, R.; Ulmer-Tuffigo, H. Observation of Bottleneck Effects on the Photoluminescence from Polaritons in II - VI Microcavities. *Phys. B* **1999**, *272*, 476–479.

(26) Senellart, P.; Bloch, J. Nonlinear Emission of Microcavity Polaritons in the Low Density Regime. *Phys. Rev. Lett.* **1999**, *82* (6), 1233–1236.

(27) Pau, S.; Cao, H.; Jacobson, J.; Björk, G.; Yamamoto, Y.; Imamoğlu, A. Observation of a Laserlike Transition in a Microcavity Exciton Polariton System. *Phys. Rev. A* **1996**, *54* (3), R1789–R1792.

(28) Imamoğlu, A.; Ram, R.; Pau, S.; Yamamoto, Y. Nonequilibrium Condensates and Lasers without Inversion: Exciton-Polariton Lasers. *Phys. Rev. A* **1996**, *53* (6), 4250–4253.

(29) Maragkou, M.; Grundy, A. J. D.; Ostatnick, T.; Lagoudakis, P. G. Longitudinal Optical Phonon Assisted Polariton Laser. *Appl. Phys. Lett.* **2010**, *97* (11), 111110.

(30) Sebald, K.; Seyfried, M.; Klembt, S.; Bley, S.; Rosenauer, A.; Hommel, D.; Kruse, C. Strong Coupling in Monolithic Microcavities with ZnSe Quantum Wells. *Appl. Phys. Lett.* **2012**, *100* (16), 161104.

(31) Klembt, S.; Durupt, E.; Datta, S.; Klein, T.; Baas, A.; Léger, Y.; Kruse, C.; Hommel, D.; Minguzzi, A.; Richard, M. Exciton-Polariton Gas as a Nonequilibrium Coolant. *Phys. Rev. Lett.* **2015**, *114* (18), 186403.

(32) Zhang, J.; Li, D.; Chen, R.; Xiong, Q. Laser Cooling of a Semiconductor by 40 Kelvin. *Nature* **2013**, *493* (7433), 504–508.

(33) Bloor, D.; Cross, G.; Sharma, P. K.; Elliott, J. A.; Rumbles, G. Frequency Up-Conversion in Fluid and Solid Solutions of the Oxazine Dye, Rhodamine B. *J. Chem. Soc. Faraday Trans.* **1993**, *89* (22), 4013–4015.

(34) Clark, J. L.; Rumbles, G. Laser Cooling in the Condensed Phase by Frequency Up-Conversion. *Phys. Rev. Lett.* **1996**, *76* (12), 2037–2040.

(35) Herrera, F.; Spano, F. C. Cavity-Controlled Chemistry in Molecular Ensembles. *Phys. Rev. Lett.* **2016**, *116* (23), 238301.

(36) Spano, F. C. Optical Microcavities Enhance the Exciton Coherence Length and Eliminate Vibronic Coupling in J-Aggregates. *J. Chem. Phys.* **2015**, *142*, 184707.

(37) Galego, J.; Garcia-vidal, F. J.; Feist, J. Cavity-Induced Modifications of Molecular Structure in the Strong-Coupling Regime. *Phys. Rev. X* **2015**, *5* (4), 041022.

(38) Cwik, J. A.; Reja, S.; Littlewood, P. B.; Keeling, J. Polariton Condensation with Saturable Molecules Dressed by Vibrational Modes. *EPL* **2014**, *105*, 47009.

(39) Zeb, M. A.; Kirton, P. G.; Keeling, J. Exact States and Spectra of Vibrationally Dressed Polaritons. *ACS Photonics* **2018**, *5* (1), 249–257.

(40) Herrera, F.; Spano, F. C. Dark Vibronic Polaritons and the Spectroscopy of Organic Microcavities. *Phys. Rev. Lett.* **2017**, *118* (22), 223601.

(41) Cortese, E.; Lagoudakis, P. G.; De Liberato, S. Collective Optomechanical Effects in Cavity Quantum Electrodynamics. *Phys. Rev. Lett.* **2017**, *119* (4), 043604.

(42) Chen, Y.; Zhao, J.; Guo, H.; Xie, L. Geometry Relaxation-Induced Large Stokes Shift in Red-Emitting Borondipyrromethenes (BODIPY) and Applications in Fluorescent Thiol Probes. *J. Org. Chem.* **2012**, *77* (5), 2192–2206.

(43) Musser, A. J.; Rajendran, S. K.; Georgiou, K.; Gai, L.; Grant, R. T.; Shen, Z.; Cavazzini, M.; Ruseckas, A.; Turnbull, G. A.; Samuel, I. D. W.; Clark, J.; Lidzey, D. G. Intermolecular States in Organic Dye Dispersions: Excimers vs. Aggregates. *J. Mater. Chem. C* **2017**, *5* (33), 8380–8389.

(44) Grant, R. T.; Michetti, P.; Musser, A. J.; Gregoire, P.; Virgili, T.; Vella, E.; Cavazzini, M.; Georgiou, K.; Galeotti, F.; Clark, C.; Clark, J.; Silva, C.; Lidzey, D. G. Efficient Radiative Pumping of Polaritons in a Strongly Coupled Microcavity by a Fluorescent Molecular Dye. *Adv. Opt. Mater.* **2016**, *4* (10), 1615–1623.

(45) Savona, V.; Andreani, L. C.; Schwendimann, P.; Quattropani, A. Quantum Well Excitons in Semiconductor Microcavities: Unified Treatment of Weak and Strong Coupling Regimes. *Solid State Commun.* **1995**, *93* (9), 733–739.

(46) Shalabney, A.; George, J.; Hutchison, J.; Pupillo, G.; Genet, C.; Ebbesen, T. W. Coherent Coupling of Molecular Resonators with a Microcavity Mode. *Nat. Commun.* **2015**, *6*, 5981.

(47) George, J.; Shalabney, A.; Hutchison, J. A.; Genet, C.; Ebbesen, T. W. Liquid-Phase Vibrational Strong Coupling. *J. Phys. Chem. Lett.* **2015**, *6* (6), 1027–1031.

(48) Long, J. P.; Simpkins, B. S. Coherent Coupling between a Molecular Vibration and Fabry-Perot Optical Cavity to Give Hybridized States in the Strong Coupling Limit. *ACS Photonics* **2015**, *2* (1), 130–136.

(49) del Pino, J.; Feist, J.; Garcia-Vidal, F. J. Quantum Theory of Collective Strong Coupling of Molecular Vibrations with a Microcavity Mode. *New J. Phys.* **2015**, *17*, 053040.

**Figures**

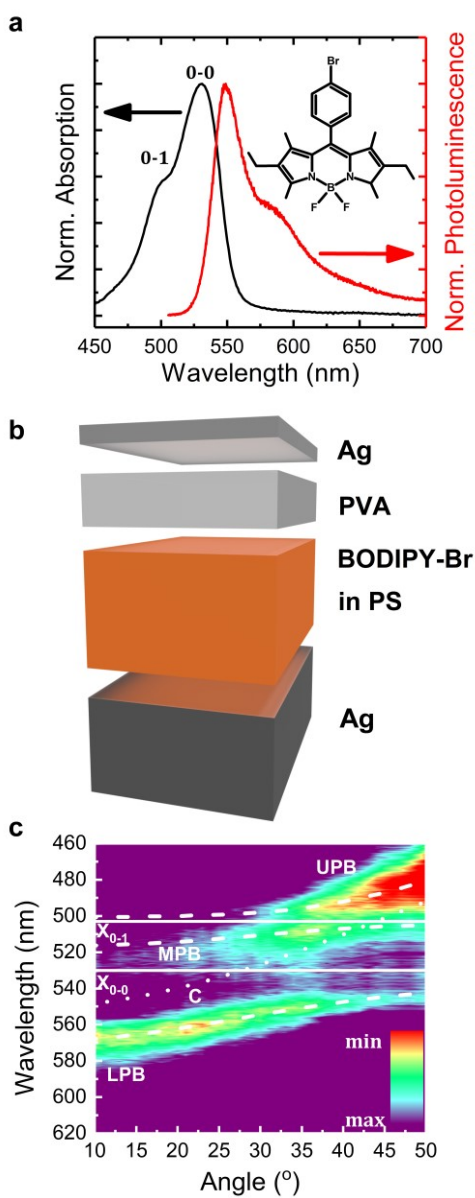

**Figure 1.** (a) Normalised absorption and photoluminescence of a control film along with the chemical structure of BODIPY-Br. (b) A schematic of the metal/metal mirror microcavities. A layer of PVA is used to protect the organic dye from the deposition of the top metal mirror and adjust the thickness of the microcavity. (c) Angular resolved white light reflectivity measurements of the microcavity recorded at room temperature. Included on the plot are the results of a three-level coupled oscillator model. LPB, MPB and UPB: Lower, middle and upper polariton branches. C, $X_{0-0}$ and $X_{0-1}$: Cavity mode, exciton from the 0-0 transition and exciton from the 0-1 vibrational transition.

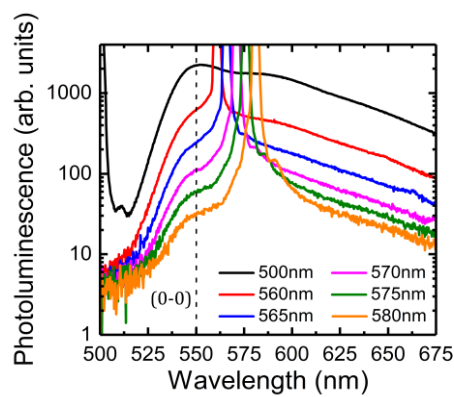

**Figure 2.** Photoluminescence and frequency up-conversion of a BODIPY-Br / PS non-cavity thin film for high direct absorption (500 nm) and sub band-gap (560 nm – 580 nm) excitation wavelength. The dashed line shows the peak emission wavelength (550 nm) of the $X_{0-0}$ electronic transition.

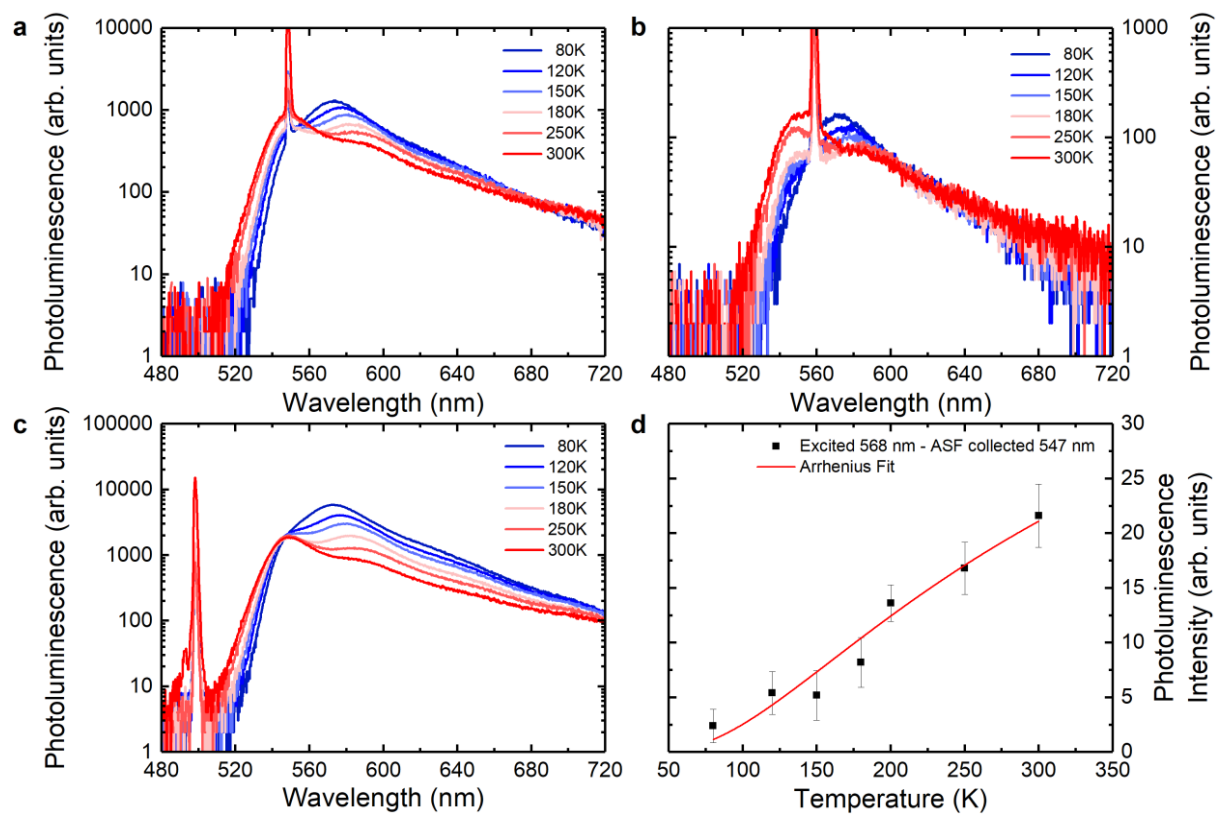

**Figure 3.** Temperature dependent photoluminescence and frequency up-conversion of a BODIPY-Br / PS non-cavity thin film following laser excitation at (a) 548 nm, (b) 558 nm and (c) 500 nm. (d) Temperature dependence of ASF intensity of a non-cavity BODIPY-Br / PS thin film when excited at 568 nm and collected at 547 nm along with an Arrhenius fit with an activation energy of around 27 meV.

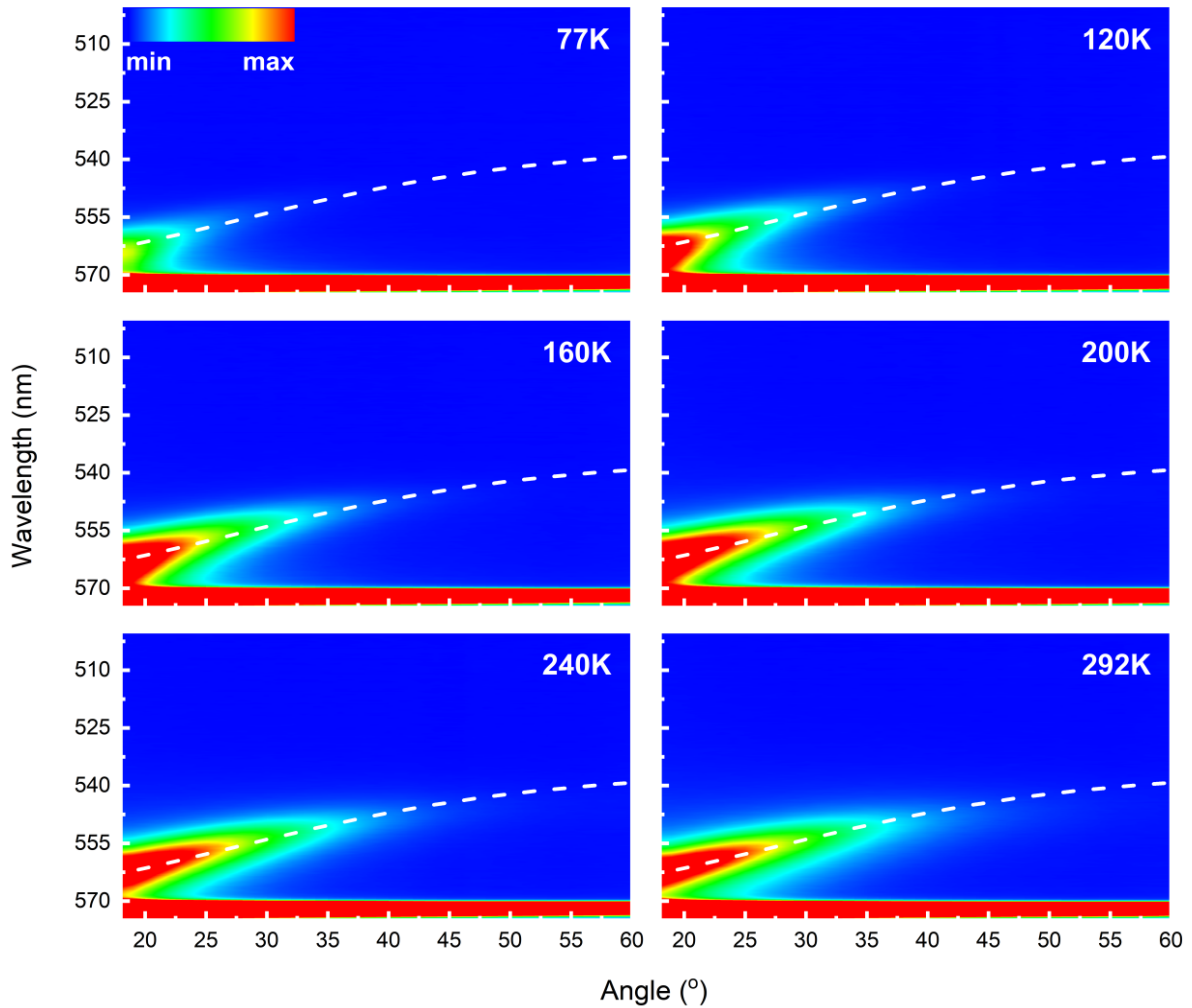

**Figure 4.** Temperature dependent photoluminescence of the lower polariton branch following resonant excitation at 572 nm showing the anti-Stokes fluorescence vs emission angle. The dashed lines indicate the energy of the lower polariton branch as derived from the white light reflectivity measurements of the microcavity using a three-level coupled oscillator model. The normalised false colour scale was saturated at around 0.7 to improve the visibility of the lower polariton branch emission.

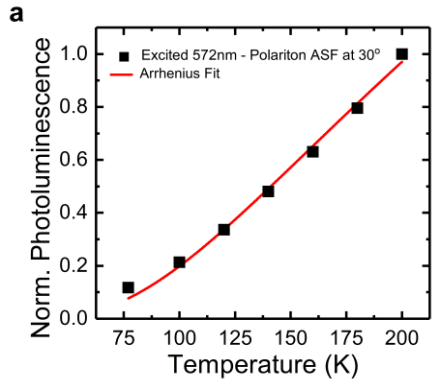

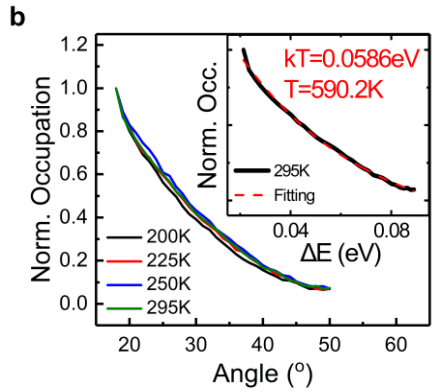

**Figure 5.** (a) Normalized intensity of the lower polariton branch emission following resonant excitation at 572 nm, recorded at an angle of 30º (555 nm), for temperatures between 77 and 200 K along with an Arrhenius fit. (b) Normalised occupation of the lower polariton branch following resonant (572 nm) excitation for four different temperatures. The inset shows the Boltzmann fitting of the room temperature lower polariton branch occupation.

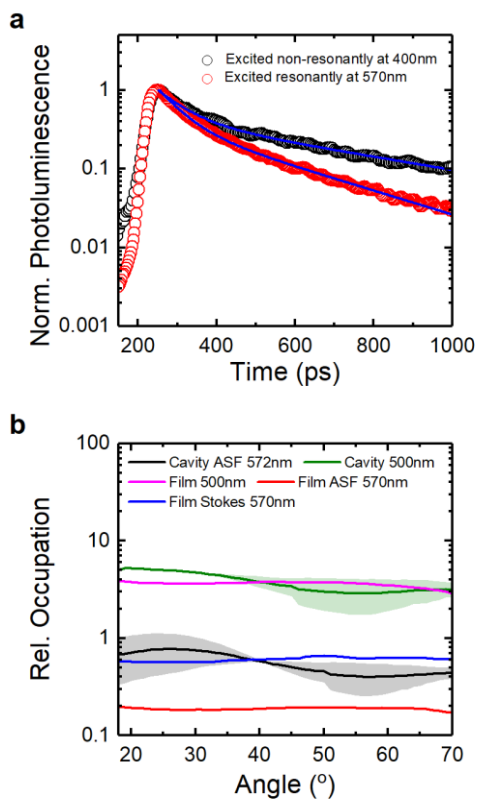

**Figure 6.** (a) Transient dynamics of the lower polariton branch emission at an angle of 28° following resonant (570 nm) and non-resonant (400 nm) excitation. (b) Lower polariton branch relative occupation following resonant (572 nm, black line) and non-resonant (500 nm, green line) excitation as a function of emission angle. Angular-resolved anti-Stokes (red line) and Stokes (blue line) fluorescence from a control film excited at 570 nm as well as angular-resolved fluorescence (magenta line) excited at 500 nm. The coloured shaded areas indicate the uncertainty of the measurement.

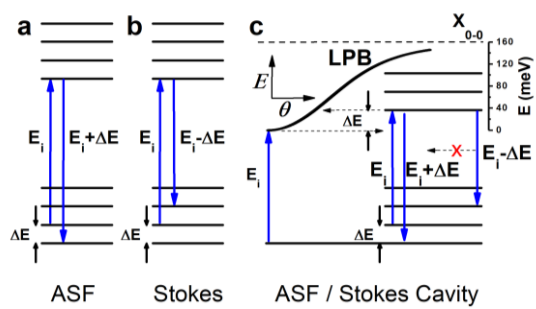

**Figure 7.** Schematic of (a) anti-Stokes fluorescence (ASF) and (b) Stokes fluorescence in a control film as well as (c) in a microcavity. Here, part (c) indicates the proposed mechanism in which anti-Stokes fluorescence is generated following resonant excitation of the bottom of the lower polariton branch in a strongly coupled organic microcavity.

# For Table of Contents Use Only:

**Generation of Anti-Stokes Fluorescence in a Strongly Coupled Organic Semiconductor Microcavity**

Kyriacos Georgiou, Rahul Jayaprakash, Alexis Askitopoulos, David M. Coles, Pavlos G. Lagoudakis and David G. Lidzey

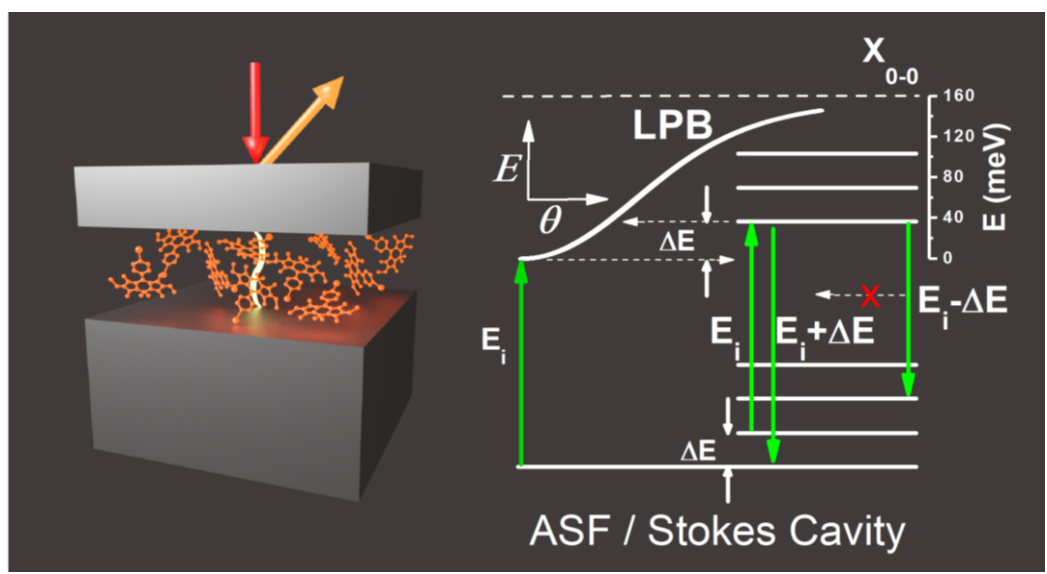

ToC: The strongly-coupled microcavity containing the fluorescent dye BODIPY-Br is shown on the left hand side. The microcavity is excited at normal incidence with a laser (red) and photoluminescence (PL) emission of a shorter wavelength (orange) is detected at different angles. This is evidence of anti-Stokes fluorescence (ASF) of organic exciton-polaritons in a strongly-coupled microcavity. On the right hand side we present the proposed mechanism for anti-Stokes and Stokes fluorescence of polaritons.